\documentclass[final]{elsarticle}

\usepackage{lineno,hyperref}
\usepackage{multirow}
\usepackage{graphicx}
\usepackage[a4paper]{geometry}
\usepackage{amssymb}

\usepackage{tabularx}
\makeatletter
\def\hlinewd#1{%
\noalign{\ifnum0=`}\fi\hrule \@height #1 %
\futurelet\reserved@a\@xhline}
\makeatother

\journal{Journal of Crystal Growth}

\begin{document}

\begin{frontmatter}

\title{Thermal analysis and crystal growth of doped Nb$_2$O$_5$}

\author[mainaddress,secondaryaddress]{Julia Hidde\fnref{fn1}}
\fntext[fn1]{This work is based on a thesis submitted by JH to the Department of Chemistry, Humboldt Universit\"at zu Berlin in partial fulfillment of the requirements for the degree of Master of Science.}
\author[mainaddress]{Christo Guguschev}
\author[mainaddress,secondaryaddress]{Detlef Klimm\corref{correspondingauthor}}

\cortext[correspondingauthor]{Corresponding author}
\ead{detlef.klimm@ikz-berlin.de}

\address[mainaddress]{Leibniz-Institut f\"ur Kristallz\"uchtung, Max-Born-Str. 2, 12489 Berlin, Germany}
\address[secondaryaddress]{Humboldt-Universit\"at zu Berlin, Institut f\"ur Chemie, Brook-Taylor-Str. 2, 12489 Berlin}

\begin{abstract}

The systems Nb$_2$O$_5$--Ta$_2$O$_5$ and Nb$_2$O$_5$--V$_2$O$_5$ were investigated using thermal analysis, X-ray powder diffraction and thermodynamic simulations. Solid solution formation is possible for both systems; furthermore, both contain one intermediate compound, VNb$_9$O$_{25}$ or Ta$_2$Nb$_4$O$_{15}$, respectively. Phase relationships for pure niobium(V)-oxide and tantalum(V)-oxide were studied under ambient pressure. It was found that both compounds can occur in two stable solid modifications. For niobium(V)-oxide this are the monoclinic high-temperature modification (H-Nb$_2$O$_5$) and an orthorhombic low-temperature modification (T-Nb$_2$O$_5$) and for tantalum(V)-oxide a tetragonal high-temperature form ($\alpha$-Ta$_2$O$_5$) and an orthorhombic low-temperature form ($\beta$-Ta$_2$O$_5$). Based on these results, crystal growth experiments with various compositions from both systems were carried out using the optical floating zone (OZF) technique.

\end{abstract}


\end{frontmatter}


\section{Introduction}

The oxides of niobium form a complex system consisting of numerous polymorphic phases. Being able to adopt different structures, niobium oxides possess interesting properties, which allow their application in science and technology. Besides the usage as solid electrolyte capacitor or complementary metal oxide semiconductors, the photochromic switching of niobium(V)-oxide coupled with the ability to transform into a conductive suboxide offers the potential to use these materials as photoelectrodes in solar cells \cite{Yon01, Ver13, Kau01, Vie10, Tak11}.

Another field of current research is the resistive switching behavior of Nb$_2$O$_5$, which was first described in 1965 by Hiatt and Hickmott \cite{Hia65}. This property of niobium(V)-oxide could allow the construction of memristors and their application as non-volatile memory devices \cite {Mah12, Gri13, Nic17}. For the production of functioning memristors, a detailed knowledge of the structural changes and other processes will be necessary, that take place during the setting of the resistance. To investigate this issue, the study of homoepitaxial niobium-oxide layers seems promising. However, substrate single crystals of niobium(V)-oxide are not commercially available.

The variety of different modifications that can occur for niobium(V)-oxide makes bulk crystal growth rather challenging. Nevertheless, Shindo and Komatsu \cite{Shindo76} succeeded in growing undoped H-Nb$_2$O$_5$ crystals with diameters of about 6\,mm and lengths of several centimeters. In the literature, contradictory information can be found on the structure and stability of the various modifications. While Orr and King \cite{Orr53,King54} describe a single monoclinic modification, which exists from room temperature to the melting point without performing a phase transition, Brauer and many other authors observe several transitions between different modifications \cite{Brau41, Schef66, Tam74, Kat75, Mes01, Pil17}.

In this paper the two systems Nb$_2$O$_5$--V$_2$O$_5$ and Nb$_2$O$_5$--Ta$_2$O$_5$ are investigated using different techniques of thermal analysis and X-ray powder diffraction, supported by a thermodynamic assessment of both systems. The aim of this study is to clarify phase equilibria of the pure oxides, as well as the two pseudo-binary systems. Additionally, an attempt was made to grow first doped Nb$_2$O$_5$-crystals using the optical floating zone technique. In this context it is of particular interest whether mixed-crystal formation of niobium(V)-oxide with the two oxides Ta$_2$O$_5$ or V$_2$O$_5$ is possible and whether single-crystal growth of niobium(V)-oxide can be stabilized by doping with the oxides of group five of the periodic table.

\section{Experimental}

For the preparation of all samples, as well as for the feed-rods and seed-rods for crystal growth, the following powdered chemicals were used: Nb$_2$O$_5$ (H.C. Starck, purity 99.995\%), Ta$_2$O$_5$ (Fox-Chemicals GmbH, 99.99\%) and V$_2$O$_5$ (Aran Isles Chemicals Inc., 99.99\%).

Pure oxides and mixtures of different composition from the systems Nb$_2$O$_5$--Ta$_2$O$_5$ and Nb$_2$O$_5$--V$_2$O$_5$ were investigated by differential thermal analysis (DTA) with simultaneous thermogravimetry (TG). For this purpose, the chemicals were weighed in varying proportions, mixed and analyzed using a NETZSCH STA 449C Jupiter thermal analyzer. A DTA/TG sample carrier with Pt/Pt90Rh10 thermocouples and platinum crucibles allowed measurements up to $1650^{\,\circ}$C in a flowing mixture of 20\,ml/min argon and 20\,ml/min oxygen. Only for the determination of the melting point of pure tantalum(V)-oxide high-temperature DTA/TG measurements were performed in a NETZSCH STA 429C analyzer with W/Re thermocouples and tungsten crucibles in static helium atmosphere.

Each DTA/TG measurement consisted of three heating/cooling cycles with rates of 10\,K/min. Between the individual heating segments, the samples were cooled to $40^{\,\circ}$C. The maximum temperature for each sample was chosen in such a way that the liquidus temperature was surpassed. The first heating was used to homogenize the sample by melting the substance mixture. Most signals for the evaluation of the measurements were taken from the second heating cycle, while the third heating segment was used to assess the reproducibility of the signals. In all measurements no significant mass changes were detected, so that a reduction of the investigated substances by oxygen release during the heating phase can be excluded.

The different crystal phases and modifications were characterized by X-ray powder diffraction with an XRD 3003 TT (GE Inspection Technologies), which has Bragg-Brentano geometry with scintillation detector. These measurements were carried out using the Cu K$_{\alpha 1}$ radiation with a wavelength of $\lambda = 154.06$\,pm. For that purpose, powder samples of the pure oxides and some mixtures from the binary systems Nb$_2$O$_5$--Ta$_2$O$_5$ and Nb$_2$O$_5$--V$_2$O$_5$ were heated to various temperatures with a heating rate of 10\,K/min and sintered for several hours using NETZSCH STA 449C Jupiter and NETZSCH STA 409C furnaces and Pt crucibles (for the exact conditions see table~\ref{Tab_Sin}). The samples were then ground in a mortar and analyzed by X-ray powder diffraction.

\begin{table}[ht]
\caption{Composition and sintering conditions of the samples investigated with XRD (holding temperature and holding time as well as cooling rate). All samples were heated at a rate of 10\,K/min. In addition, the untreated oxides were used to investigate the room temperature modifications. These samples are referred to as Nb1 (untreated Nb$_2$O$_5$), Ta1 (Ta$_2$O$_5$) and V1 (V$_2$O$_5$).}
\begin{tabular}{lccccc}
\hline
Name   & Substance                 & $x$(Nb$_2$O$_5$) & Temp. [$^\circ$C] & Time [h] & Cooling [K/min] \\ \hline
Nb2    & Nb$_2$O$_5$               & 1                & 720               & 50       & 10 \\
       &                           &                  & 550               & 50       & 10 \\ 
Nb3    & Nb$_2$O$_5$               & 1                & 1460              & 20       & 30 \\ 
Ta2    & Ta$_2$O$_5$               & 0                & 1460              & 20       & 30 \\ 
Ta3    & Ta$_2$O$_5$               & 0                & 1550              & 15       & 60 \\ 
V2     & V$_2$O$_5$                & 0                & 600               & 20       & 40 \\ \hline
NbV1   & Nb$_2$O$_5$ + V$_2$O$_5$  & 0.96             & 700               & 20       & 30 \\ 
NbV2   & Nb$_2$O$_5$ + V$_2$O$_5$  & 0.88             & 700               & 20       & 30 \\ 
NbTa1  & Nb$_2$O$_5$ + Ta$_2$O$_5$ & 0.95             & 1460              & 20       & 30 \\ 
NbTa2  & Nb$_2$O$_5$ + Ta$_2$O$_5$ & 0.86             & 1460              & 20       & 30 \\ \hline
\end{tabular}
\label{Tab_Sin}
\end{table}

Crystal growth experiments of doped niobium(V)-oxide were carried out in an optical floating zone (OFZ) furnace FZ-T-10000-H-VII-VPO-PC (Crystal System Corporation) with four ellipsoidal mirrors and four 1500\,W halogen bulbs. Best results were achieved using a pre-molten feed-rod, a growth rate of 6\,mm/h and a rotation rate of 30\,rpm. Flowing dried air (2~l/min) under ambient pressure was the growth atmosphere.

\section{Results and Discussion}

\subsection{Phases of the pure oxides}
\label{sec:pure}

\begin{figure}[ht]
\centering
\includegraphics[width=0.5\textwidth]{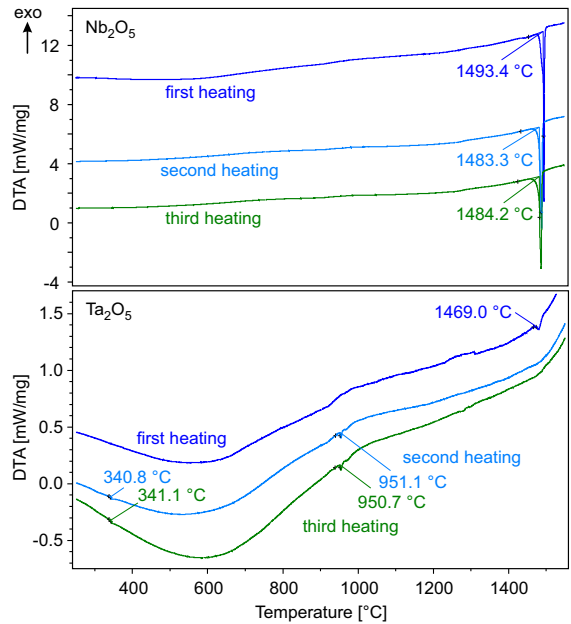}
\caption{Typical DTA heating curves of a complete measurement cycle with three heating segments from Nb$_2$O$_5$ and Ta$_2$O$_5$. The temperature at which a phase transition occurs can be determined as the onset of the signals.}
\label{Fig_DTA}
\end{figure}

Fig.~\ref{Fig_DTA} shows DTA curves of three subsequent heating segments with Nb$_2$O$_5$ or Ta$_2$O$_5$, respectively. For Nb$_2$O$_5$, only the melting point $T_\mathrm{f}=(1484\pm4)^{\,\circ}$C is obvious as a sharp, endo\-thermic peak; no further signals are visible during heating and cooling. The somewhat higher onset of the melting peak during the first heating might result from a loose packing of the powder sample with consequently poor thermal contact to the crucible wall. Literature data on the melting point of Nb$_2$O$_5$ are largely inconsistent: values ​​of $1465^{\,\circ}$C \cite{Schef58}, $1479^{\,\circ}$C \cite{Scha91}, $1491^{\,\circ}$C \cite{Hol56} and $1512^{\,\circ}$C \cite{FactSage} are published. The melting point observed in this work is within the range defined by the literature.

\begin{figure}[ht]
\centering
\includegraphics[width=0.5\textwidth]{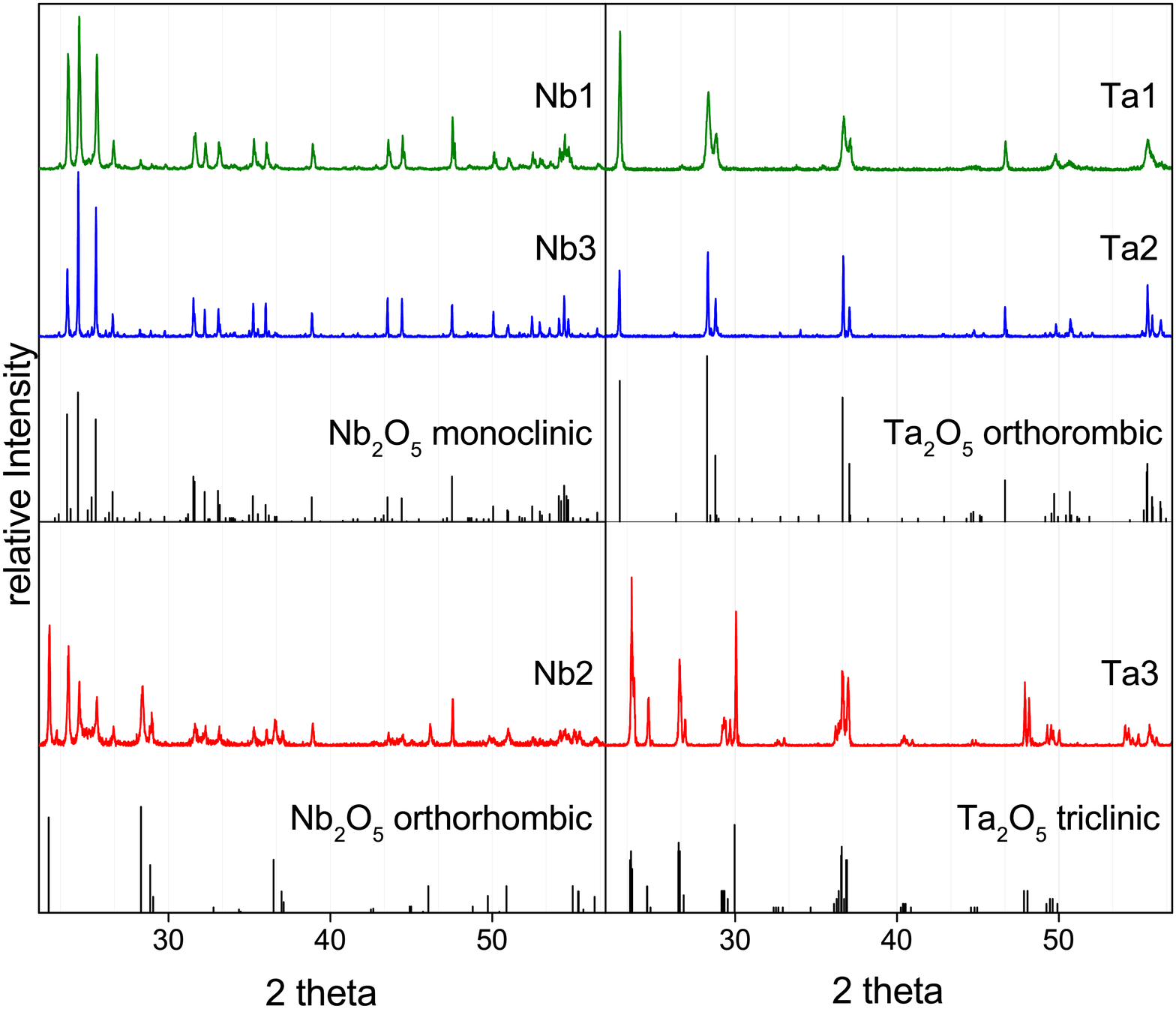}
\caption{Left: X-ray diffractograms of Nb$_2$O$_5$ compared to literature data (black curves) for monoclinic H-Nb$_2$O$_5$ \cite{PDF37} and orthorhombic T-Nb$_2$O$_5$ \cite{PDF27}. By sintering at temperatures around $700^{\,\circ}$C, it was possible to achieve the transition to the low-temperature modification. Right: X-ray diffractograms of Ta$_2$O$_5$ compared to literature data for orthorhombic $\beta$-Ta$_2$O$_5$ \cite{PDF25} and triclinic $\alpha''$-Ta$_2$O$_5$ \cite {PDF21}, which is a metastable low-temperature form of the tetragonal $\alpha$-Ta$_2$O$_5$ \cite{PDF21-99}.}
\label{Fig_XRD}
\end{figure}

In contrast to the DTA measurements, X-ray powder diffraction of pure niobium(V)-oxide reveals the presence of two modifications. As shown in Fig.~\ref{Fig_XRD} (left side), all reflections observed in the diffractograms of samples Nb1 and Nb3 can be fully explained by a monoclinic phase, which is consistent with the high-temperature modification H-Nb$_2$O$_5$ \cite{Brau41, PDF37}. Sample Nb2 has additional signals that can be assigned to an orthorhombic low-temperature modification T-Nb$_2$O$_5$ \cite{Brau41, PDF27}. By long-term sintering for 50~hours at temperatures around $700^{\,\circ}$C, it was apparently possible to achieve partial transition to this low-temperature modification, whereas the raw sample Nb1 consists of the H-Nb$_2$O$_5$ phase which is obviously metastable.

Contrary to the current work, conversion from H-Nb$_2$O$_5$ to T-Nb$_2$O$_5$ could not be observed by Brauer. He described that even after 12\,days of sintering at $850^{\,\circ}$C only the pure H-modification was present \cite{Brau41}, which excludes a transition beyond this temperature. Also Holtzberg et al. and Waring et al. found no reversibility in various annealing experiments \cite{Hol56, War73}. Therefore, Waring concluded that the T-modification was a metastable state. This conclusion cannot be supported in the present work. Rather, it is assumed that the transformation is a kinetically inhibited reaction, so that the thermodynamic minimum can only be reached by long-term sintering.

The DTA curves of tantalum(V)-oxide in Fig.~\ref{Fig_DTA} (bottom) show several phase transitions up to $1640^{\,\circ}$C. During first heating, a single endothermic signal appears at $1469^{\,\circ}$C, while other endothermic signals at $341^{\,\circ}$C and $951^{\,\circ}$C appear during the following heating segments. The melting point is beyond the temperature range of this measurement, but measurements in the STA 429 showed $T_\mathrm{f}=(1883\pm20)^{\,\circ}$C (Fig.~\ref{Supp_Ta-HT} in the supplementary material), which is in good agreement with literature data $T_\mathrm{f}=1875^{\,\circ}$C \cite{Ruf13} and $1878^{\,\circ}$C \cite{FactSage}.

Furthermore, the X-ray powder diffraction measurements of tantalum(V)-oxide reveal the presence of two modifications (see Fig.~\ref{Fig_XRD}). While the diffractograms of the samples Ta1 and Ta2 show only the orthorhombic low-temperature modification \cite{PDF25}, the X-ray diffractogram of sample Ta3 differs significantly from the previous ones. The signals of this diffractogram can be assigned to a triclinic crystal structure \cite{PDF21}. It should be noted that only the annealing temperature of Ta3 (Tab.~\ref{Tab_Sin}) is beyond the phase transition that was detected during the first heating of Ta$_2$O$_5$ (Fig.~\ref{Fig_DTA}).

According to the literature, tantalum oxide exists in two polymorphic modifications $\alpha$-Ta$_2$O$_5$ (tetragonal high-temperature form) and $\beta$-Ta$_2$O$_5$ (orthorhombic low-temperature form). However, there is some disagreement in the phase transition temperature. Lagergren et al. report the existence of a high-temperature modification above $T_\mathrm{t}=1320^{\,\circ}$C \cite{Lag52}, while Waring et al. observe this modification only above $1360^{\,\circ}$C \cite{War68}. F\"uglein et al. found the transition from $\beta$-Ta$_2$O$_5$ to $\alpha$-Ta$_2$O$_5$ at $1482^{\,\circ}$C \cite{Fug02}, which is in fairly good agreement with the DTA results from this work. In addition to these two thermodynamically stable modifications, tantalum(V)-oxide has several metastable structures \cite{War68, Fug02, Hum92, Garg96}. Waring et al. observe a tetragonal modification, which on cooling first changes to a monoclinic and then to a triclinic modification \cite{War68}. Recently, the same results were obtained by F\"uglein et al. \cite{Fug02}.

Based on the observations of Waring and F\"uglein, the present work assumes that $\alpha$-Ta$_2$O$_5$ and $\beta$-Ta$_2$O$_5$ are the only thermodynamically stable solid modifications, with $T_\mathrm{t}=1469^{\,\circ}$C, as shown during the first heating of Ta$_2$O$_5$ in Fig.~\ref{Fig_DTA}. No indications for the back transformation $\alpha\rightarrow\beta$ were found in the DTA cooling curves, which indicates that the high-$T$ form can exist metastable also below $T_\mathrm{t}$. However, this metastable $\alpha$-Ta$_2$O$_5$ tends to transform to the metastable modifications $\alpha'$ and $\alpha''$ upon cooling. The room temperature metastable $\alpha''$-form has then a triclinic crystal structure, which is in agreement with the results of the X-ray powder diffraction spectrum of sample Ta3 in Fig.~\ref{Fig_XRD}. The DTA peaks at $341^{\,\circ}$C and $951^{\,\circ}$C (Fig.~\ref{Fig_DTA}) results from transformations between the metastable modifications of Ta$_2$O$_5$.

Consistent with the literature, for V$_2$O$_5$ only one stable modification was found by X-ray powder diffraction. Also, the DTA measurements of vanadium(V)-oxide show only one sharp, endothermic signal at $668^{\,\circ}$C, which can be attributed to the melting point (Fig.~\ref{Supp_V2O5}). This value is in very good agreement with the literature where $T_\mathrm{f}=669.8^{\,\circ}$C is given \cite{FactSage}.

\subsection{The binary system Nb$_2$O$_5$--V$_2$O$_5$}
\label{sec:Nb-V}

\begin{figure}[ht]
\centering
\includegraphics[width=0.5\textwidth]{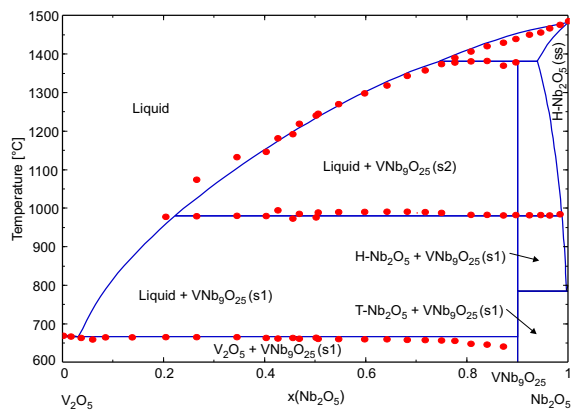}
\caption{DTA signals (dots) and thermodynamic assessment of the system Nb$_2$O$_5$--V$_2$O$_5$.}
\label{Fig_Phase1}
\end{figure}

The system Nb$_2$O$_5$--V$_2$O$_5$ was first investigated by Waring and Roth in 1964 \cite{War65}. Basic features of their phase diagram could be confirmed by the current results that are shown in Fig.~\ref{Fig_Phase1}. This diagram is set up by the components V$_2$O$_5$, Nb$_2$O$_5$ (H- and T-phase) and by an intermediate compound VNb$_9$O$_{25}$. In addition to the previous report \cite{War65}, a transition between low- and high-$T$ phases s1 and s2 of VNb$_9$O$_{25}$ was found in the current study, leading to strong DTA peaks near $990^{\,\circ}$C. Other DTA peaks that occur at constant temperatures over a wide concentration range result from the eutectic close to V$_2$O$_5$ ($T_\mathrm{eut}=663^{\,\circ}$C) and from the peritectic decomposition of VNb$_9$O$_{25}$ at $T_\mathrm{per}=1375^{\,\circ}$C. This peak appears only for compositions $0.75\leq x($Nb$_2$O$_5)\leq0.9$; the lower limit defines the peritectic point, and the upper limit defines the solubility limit of V$_2$O$_5$ in the H-Nb$_2$O$_5$(ss) solid solution phase.

Waring and Roth describe a larger series of intermediate compounds, e.g. V$_2$Nb$_{23}$O$_{62}=\frac{1}{2}$\,V$_4$Nb$_{46}$O$_{124}$ in the niobium oxide-rich part of the phase diagram \cite{War65}. These results cannot be confirmed in the present work, and only VNb$_9$O$_{25}$ was found. This observation is in agreement with other authors \cite{Scha91, Scha92, Pol94, Sur06, Tab09}. Schadow et al. were able to isolate only one stable compound by studying the system using chemical transport reactions \cite{Scha91, Scha92}. The phase diagram constructed on the basis of these results shows correspondingly great agreement with that of the present work. The additional intermediate compounds discovered by Waring and Roth actually have a lower oxygen content and are therefore not part of the examined system \cite{Tab09}. For the example mentioned above one finds e.g. V$_4$Nb$_{46}$O$_{124}=2$\,V$_2$O$_5\cdot22$\,Nb$_2$O$_5\cdot$\,Nb$_2$O$_4$.

Fig.~\ref{Fig_Phase1} summarizes DTA results of all measurements performed in the system Nb$_2$O$_5$--V$_2$O$_5$ together with a thermodynamic assessment using the software FactSage~7.1, a program for determining the equilibrium state in chemical systems from $G(T,x)$ functions for the different phases ($x$ -- molar fractions). To calculate this graph, data for the solids were taken from the FactSage databases \cite{FactSage} and modified to match the experimental observations. For V$_2$O$_5$, only the transition from the solid to the liquid phase was defined at $T_\mathrm{f}=669.8^{\,\circ}$C, while the FactSage data for Nb$_2$O$_5$ were supplemented by another solid phase such that it has two solid and one liquid phase. Since the transformation from the low-temperature phase to the high-temperature phase could not be determined in the present work, this value was set to $T_\mathrm{t}=800^{\,\circ}$C according to Shafer and Roy \cite{Schef58}. Additionally, the intermediate VNb$_9$O$_{25}=\frac{1}{2}$(V$_2$O$_5\cdot9$\,Nb$_2$O$_5)$ was introduced using weigh\-ted data from the component oxides. In order to include mixed phases in the calculations, two solutions were generated. These are on the one hand the solid solution in the high-temperature phase of Nb$_2$O$_5$ and on the other hand the melt.

A suitable model for the excess Gibbs free energy of the melt was found by Redlich-Kister polynomials \cite{Redlich48,Klimm} for a binary system of components A and B.

\begin{equation}
G^{ex} = x_A x_B \cdot \sum_{j=0}^N L_j (x_A - x_B)^j       \label{Eq_Gex}
\end{equation}

Here, $x_A$ and $x_B$ describe the mole fractions of the components and $L_j$ is the $j$-th interaction parameter of the two components. All thermodynamic data of the components used for the simulations as well as the adjusted interaction parameters are given in the tables~\ref{Supp_Par1}-\ref{Supp_Par5} in the supplementary materials.

The determined marginal solubility of V$_2$O$_5$ in Nb$_2$O$_5$ is in agreement with the work of Waring and Roth \cite{War65}; whereas Schadow et al. \cite{Scha91} found no significant formation of a solid solution. It was noted there, however, that solubility could not be accurately determined by the methods used by these authors and therefore, the lack of solubility in \cite{Scha91} is not considered to contradict the present work. The dissolution of V$_2$O$_5$ in Nb$_2$O$_5$ is confirmed also by the X-ray diffractograms of the mixture samples NbV1 and NbV2. While NbV1 shows only reflections that can be assigned to pure Nb$_2$O$_5$, all reflections occurring in the diffractogram of the sample NbV2 can be interpreted using literature data from VNb$_9$O$_{25}$ (see Figs.~\ref{Supp_XRD1} and \ref{Supp_XRD2}). Since no reflections of the intermediate are found in the sample NbV1, it can be assumed that a mixed crystal is formed and therefore Nb$_2$O$_5$ can dissolve at least 4\% V$_2$O$_5$.

\subsection{The binary system Nb$_2$O$_5$--Ta$_2$O$_5$}

Already Holtzberg et al. and Mohanty et al. \cite{Hol61, Moh64} reported the phase diagram of the system Nb$_2$O$_5$--Ta$_2$O$_5$. Both papers assumed that niobium(V)-oxide is stable in the H-modification over the entire range of the solid phase, which is in contrast to the current study (see section~\ref{sec:pure}). However, the phase diagrams differ little from the diagram determined in this work, because the H-/T-transformation of Nb$_2$O$_5$ (see previous section~\ref{sec:Nb-V}) takes place at $T_\mathrm{t}=800^{\,\circ}$C, which is below the displayed temperature range.

\begin{figure}[ht]
\centering
\includegraphics[width=0.5\textwidth]{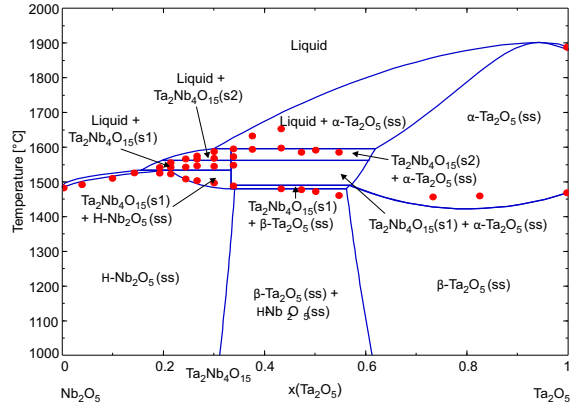}
\caption{Summary of the DTA signals of various mixtures in the system Nb$_2$O$_5$--Ta$_2$O$_5$ (dots) and thermodynamic assessment.}
\label{Fig_Phase2}
\end{figure}

Fig.~\ref{Fig_Phase2} summarizes the DTA signals of all measurements performed in the system Nb$_2$O$_5$--Ta$_2$O$_5$, where the temperatures of the detected signals are plotted against the composition of the examined sample. Again, thermodynamic calculations were carried out using FactSage~7.1 data \cite{FactSage} with Redlich-Kister polynomials. For both pure components, the thermodynamic data were supplemented with an additional solid phase. For this purpose, the $\alpha/\beta$ transition temperature $T_\mathrm{t}=1469^{\,^\circ}$C for Ta$_2$O$_5$ was used (section~\ref{sec:pure}). Besides, an intermediate compound Ta$_2$Nb$_4$O$_{15}$ was introduced. Two solid solutions between the $\beta$- or $\alpha$-modifications of Ta$_2$O$_5$ and Nb$_2$O$_5$ describe the edge solubility in the tantalum-rich part of the diagram. Furthermore, a solution of H-Nb$_2$O$_5$ and Ta$_2$O$_5$ describes the edge solubility in the niobium-rich part of the diagram. Finally, a solution of the two liquid phases was defined, which represents the melt in the investigated system (see tables~\ref{Supp_Par1}-\ref{Supp_Par5} in the supplementary materials).

The phase diagram of the system Nb$_2$O$_5$--Ta$_2$O$_5$ shows strongly pronounced edge solubilities of niobium(V)-oxide and tantalum(V)-oxide, which are interrupted only in the concentration range $0.3\lesssim x$(Ta$_2$O$_5$) $\lesssim0.6$. The intermediate compound Ta$_2$Nb$_4$O$_{15}$ is stable only in a narrow $T$-range; DTA data suggest that this compound has a phase transition s1/s2 at about $1560^{\,^\circ}$C and decomposes peritectically at $1590^{\,^\circ}$C. For Ta$_2$O$_5$ concentrations $x<0.2$, the upper $T$-limit of the H-Nb$_2$O$_5$ phase field is the solidus of this phase, and the onset of melting leads to strong DTA peaks there. For $x>0.2$, however, the phase field is limited by the beginning formation of Ta$_2$Nb$_4$O$_{15}$(s1) from the H-(Nb$_{1-x}$Ta$_x$)$_2$O$_5$ solid solution. Because in this process exclusively solid phases are involved, the enthalpy change is significantly smaller than for the melting process at $x<0.2$, and hence the corresponding DTA effect is weak. The existence of single phase H-Nb$_2$O$_5$(ss) below these phase boundaries could be confirmed by X-ray diffraction of NbTa1 ($x$(Ta$_2$O$_5)=0.05$) and NbTa2 ($x$(Ta$_2$O$_5)=0.14$), respectively (Tab.~\ref{Tab_Sin} and Fig.~\ref{Supp_XRD1}).

The recent studies \cite{Hol61, Moh64} described the limits of the solubility gap $\beta$-Ta$_2$O$_5$+H-Nb$_2$O$_5$ by vertical lines; in contrast they are described by inclined curves in the present work (Fig.~\ref{Fig_Phase2}). However, this course of the solubility limits seems more reasonable, since usually the solubility increases with $T$ \cite{Ben37}. The phase transition temperature of pure Ta$_2$O$_5$, $T_\mathrm{t}=1469^{\,\circ}$C of the current study, is more than 100\,K higher, compared to recent studies \cite{Hol61, Moh64} (see section~\ref{sec:pure}).

The intermediate compound Ta$_2$Nb$_4$O$_{15}$ was first discovered by Holtzberg et al. \cite{Hol61} and then crystallographically analyzed by Ganesh et al. \cite{Gan62}. While Holtzberg was able to observe this compound at temperatures above $1450^{\,\circ}$C, Ganesh describes that the compound could already be synthesized at temperatures of $1400^{\,\circ}$C \cite{Gan62}. Accordingly, the formation temperature of Ta$_2$Nb$_4$O$_{15}$ in the phase diagram of Mohanty et al. is $1390^{\,\circ}$C \cite{Moh64}. In the present work, this compound was not separately investigated. However, the lowermost DTA effects near $1480^{\,\circ}$C around $x$(Ta$_2$O$_5)=\frac{1}{3}$ indicate, that the stability range of Ta$_2$Nb$_4$O$_{15}$ does not extend to significantly lower temperatures.

In contrast to the work of Mohanty, in this work the phase width of the intermediate compound was disregarded. Here, a small solubility for excess Nb$_2$O$_5$ and/or Ta$_2$O$_5$ cannot be excluded. Since the phase width could not be verified in this study, no further solutions were created for this compound during the thermodynamic simulations. However a phase transformation of this compound is detected using the DTA measurements and the DTA curves show a slight, monotonous increase in the transition temperature, which could be an indication for some phase width (a few per cent) of the intermediate compound. 

Another peculiarity of the calculated phase diagram is the occurrence of an azeotropic point in the solid-liquid phase boundary in the tantalum-rich region of the diagram, resulting from the $G^{ex}(T,x)$ functions of the corresponding phases. None of the phase diagrams described in the literature contains such a point, but it can neither be refuted nor confirmed on the basis of the experimental data.

\subsection{Crystal growth of doped Nb$_2$O$_5$}

\begin{figure}[ht]
\centering
\includegraphics[width=0.5\textwidth]{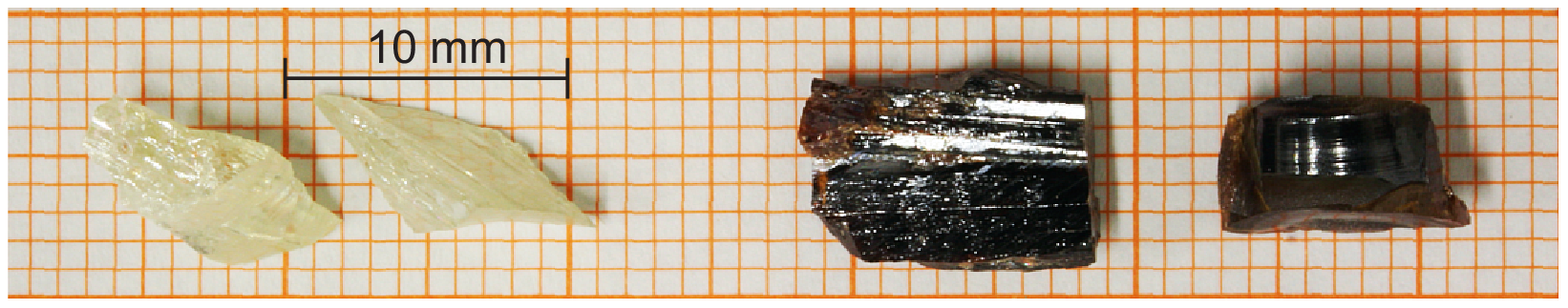}
\caption{Left: Ta:Nb$_2$O$_5$ crystals with $x($Ta$_2$O$_5) = 0.03$. Right: V:Nb$_2$O$_5$ crystals with $x($V$_2$O$_5) = 0.01$.}
\label{Fig_Cryst}
\end{figure}

The thermodynamic investigations of the systems Nb$_2$O$_5$--V$_2$O$_5$ and Nb$_2$O$_5$--Ta$_2$O$_5$ proved significant solubility of Nb$_2$O$_5$ for vanadium(V)-oxide and especially for tantalum(V)-oxide. Therefore, it should be possible to grow niobium oxide crystals that are doped with V$_2$O$_5$ or Ta$_2$O$_5$, respectively. Crystal growth experiments were carried out by optical floating zone technique, starting from feed-rods with 1--5\,mol-\% of the dopant. Within the scope of the present work it was possible to grow multicrystalline niobium(V)-oxide crystals with lengths up to 5\,cm and diameters of about 1\,cm. By cutting these crystals, single crystalline samples with lengths of about 6\,mm could be obtained for Ta:Nb$_2$O$_5$, and with lengths of 10\,mm for V:Nb$_2$O$_5$ (Fig.~\ref{Fig_Cryst}).

\begin{figure}[ht]
\centering
\includegraphics[width=0.8\textwidth]{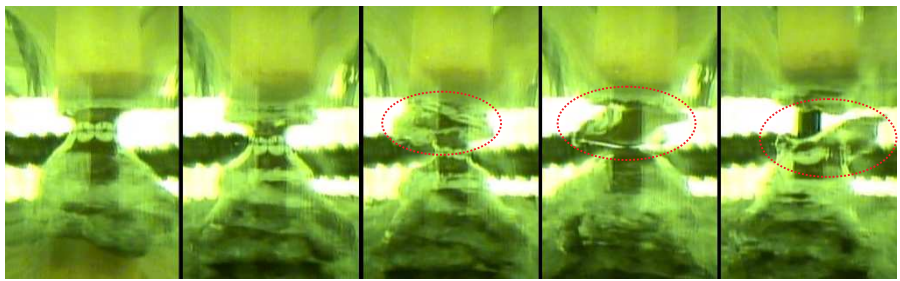}
\caption{The formation of a ``crown'' during crystal growth of doped Nb$_2$O$_5$ crystals leads to unstable growth conditions and often to the destruction of the melt zone.}
\label{Fig_Zone}
\end{figure}

The crystal growth process was typically associated with the formation of a solid ``crown'' and/or the appearance of ``noses'' at the transition from the feed rod into the molten zone. This atypical melting/solidification characteristics close to the seed resulted in a particular technical challenge, since the progressive growth of these features consumed a fraction of the melt and additionally hampered the heat input towards the residual melt. This effect appeared cyclically (at a longer time scale), and was often followed by a sudden increase in melt volume during melting of the ``crown'' or ``noses'' and sometimes resulted in demolition of the molten zone (Fig.~\ref{Fig_Zone}).

Within the scope of the present work it was not possible to eliminate the crown formation, but conditions could be found under which it was minimized. These include the use of pre-molten feed-rods and the application of higher light fluxes. In crystal growth, heat conduction plays a major role within the growing crystal and its environment \cite{Bro88, Lao16}. For continuous crystallization, stable flow of heat through the crystal is required. Additionally, the uniform melting of the starting material requires a homogeneous heat conduction within the feed-rod. Here pre-molten feed-rods seem to be superior with respect to feed-rods that were prepared by a conventional ceramic route. One possible origin of the crown formation could be an uneven heat input to various areas of the feed rod, which causes some areas to be heated more than others. Melting of the warmer regions would then lead to the formation of the observed crown structure.

\section{Summary and conclusions}

A thermodynamic assessment of the two systems Nb$_2$O$_5$--Ta$_2$O$_5$ and Nb$_2$O$_5$--V$_2$O$_5$ was carried out using thermal analysis and X-ray powder diffraction. The results showed that mixed-crystal formation is possible between Nb$_2$O$_5$ and vanadium(V)-oxide as well as Nb$_2$O$_5$ and tantalum(V)-oxide. Therefore, the phase diagrams of both systems have a marginal solubility of niobium(V)-oxide, which allows the growth of doped crystals and it was possible to obtain Ta:Nb$_2$O$_5$ as well as V:Nb$_2$O$_5$ single crystals with a size of up to 10\,mm.

Despite the partially stable crystal growth, the resulting crystals always had cracks. In order to optimize the growth conditions, the use of an afterheater or the reduction of the growth rate \cite{Shindo76} could be considered. It should be noted, however, that with a sufficiently long hold time, the kinetically inhibited phase transformation from H-Nb$_2$O$_5$ (which crystallizes first) to the low-temperature modification T-Nb$_2$O$_5$ can occur.

A particular challenge in the crystal growth of doped Nb$_2$O$_5$ is the formation of a ``crown'' at the transition from the feed-rod to the molten zone, resulting in unstable growth conditions. About the cause of this crown formation can only be speculated at the current time. Since only very few observations on such a behavior are described in the literature \cite{Wiz09}, further investigations in this area are necessary. One potential solution of this issue could be the usage of lamps with narrower filament sizes to reduce the height of the illuminated area.

\section*{Acknowledgements}

The authors gratefully thank T. Wurche for cutting the crystals, A. Kwasniewski for performing X-ray powder diffraction measurements, and M. Schwitzkowski for technical hints. D. Schulz is acknowledged for reading the manuscript.




\pagebreak
\appendix

\section{Supplementary Materials}

\setcounter{figure}{0}
\setcounter{table}{0}

\begin{figure}[ht]
\centering
\includegraphics[width=0.45\textwidth]{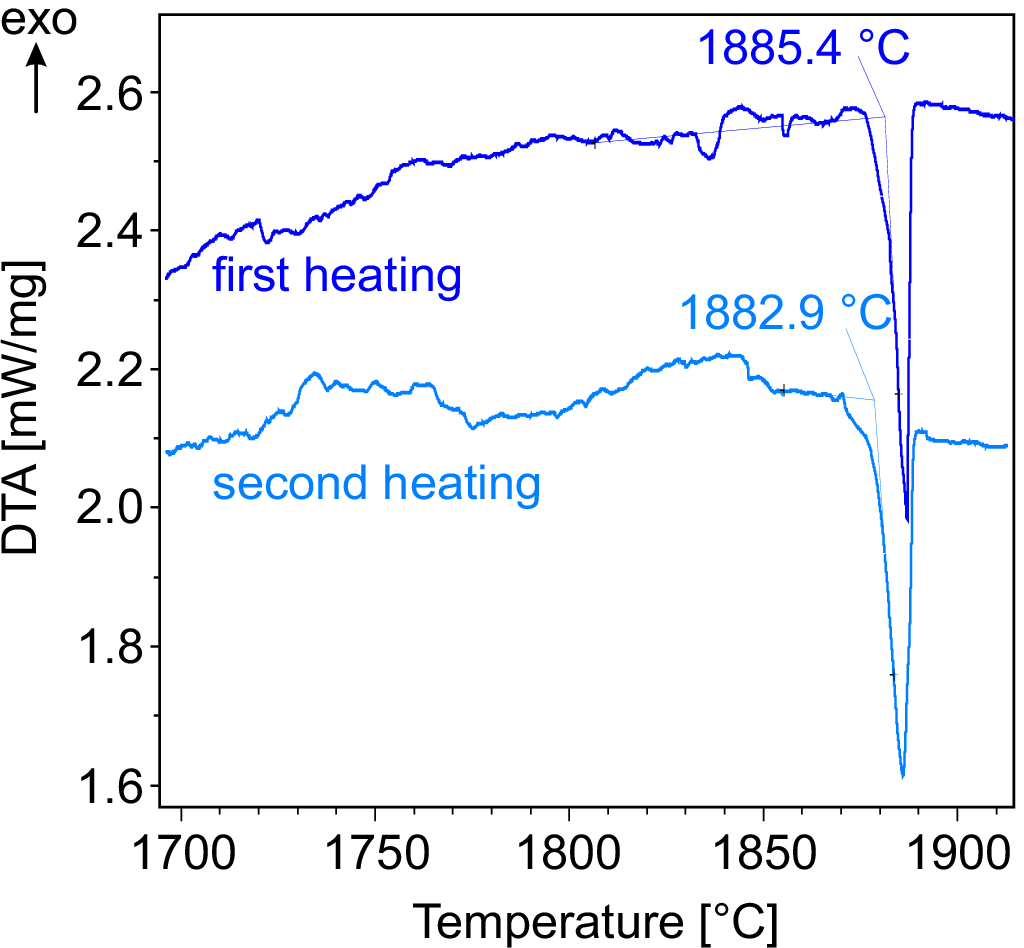}
\caption{High-temperature DTA heating curves of a measurement cycle with two heating segments for the determination of the melting point of Ta$_2$O$_5$. The temperature at which a phase transition occurs can be determined as the onset of the signals.}
\label{Supp_Ta-HT}
\end{figure}

\begin{figure}[ht]
\centering
\includegraphics[width=0.45\textwidth]{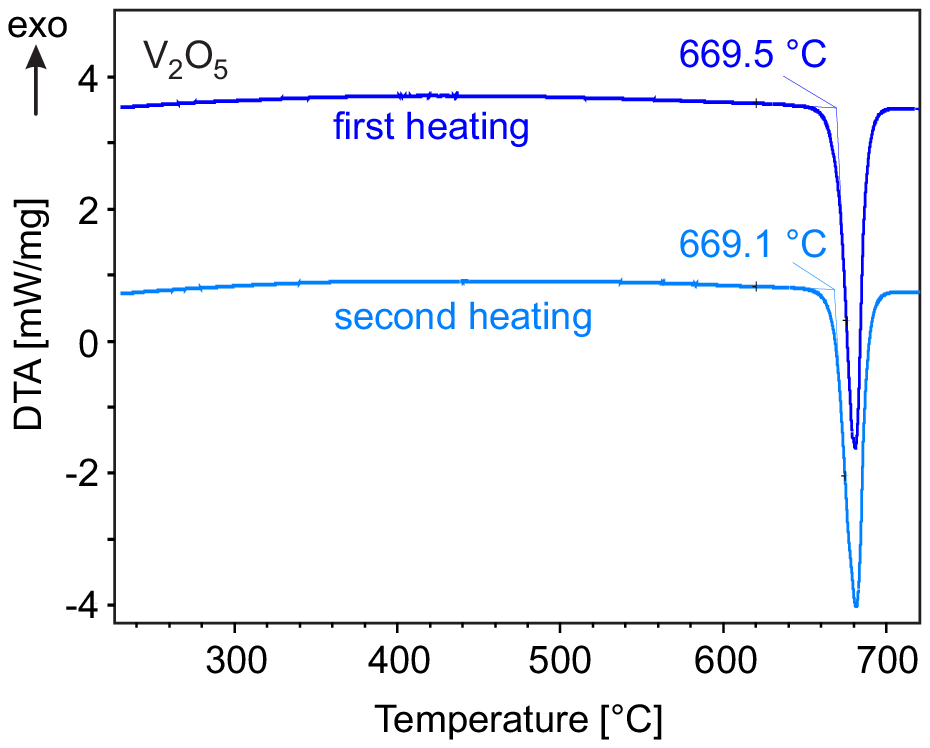}
\caption{DTA heating curves of a measurement cycle with two heating segments for the determination of the melting point of V$_2$O$_5$. The temperature at which a phase transition occurs can be determined as the onset of the signals.}
\label{Supp_V2O5}
\end{figure}

\begin{table}[ht]
\centering
\caption{Thermodynamic parameters of the FactSage simulation for the pure oxides in their solid phases S1 and S2 and their liquid Phase L1: standard enthalpy of formation $\Delta H$ and standard entropy $S$.}
\begin{tabular}{ccccccc}
\hline
   & \multicolumn{2}{c}{Ta$_2$O$_5$} & \multicolumn{2}{c}{Nb$_2$O$_5$} & \multicolumn{2}{c}{V$_2$O$_5$} \\
   & $\Delta H$ & $S$        & $\Delta H$ & $S$        & $\Delta H$ & $S$     \\
   & [J/mol]    & [J/mol\,K] & [J/mol]    & [J/mol\,K] & [J/mol]    & [J/mol\,K] \\
\hline
S1 & -2045976   & 143.114    & -1899536    & 137.298   & -1550590   & 130.549 \\
S2 & -2061517   & 111.026    & -1897536    & 139.162   & --         & -- \\
L1 & -1960862   & 178.959    & -1827201    & 175.256   & -1491202   & 191.958\\
\hline
\end{tabular}
\label{Supp_Par1}
\end{table}

\begin{table}[ht]
\centering
\caption{Thermodynamic parameters of the FactSage simulation for the intermediate compounds: standard enthalpy of formation $\Delta H$ and standard entropy $S$. S1 and S2 denote the two solid phases.}
\begin{tabular}{ccccc}
\hline
   & \multicolumn{2}{c}{V$_2$Nb$_9$O$_{25}$} & \multicolumn{2}{c}{Ta$_2$Nb$_4$O$_{15}$} \\
   & $\Delta H$ [J/mol] & $S$ [J/mol\,K] & $\Delta H$ [J/mol] & $S$ [J/mol\,K] \\
\hline
S1 & -9359107 & 694.116 & -5843789 & 401.62 \\
S2 & -9343180 & 733.616 & -5834419 & 399.09  \\
\hline
\end{tabular}
\label{Supp_Par2}
\end{table}

\begin{table}[ht]
\centering
\caption{Heat capacity (in J/mol\,K) of the intermediate compound Ta$_2$Nb$_4$O$_{15}$ as a function of temperature $c_p = a + b \cdot T + c \cdot T^{-2} + d \cdot T^2 $. The two solid phases are denoted by S1 and S2, respectively.}
\begin{tabular}{ccccc}
\hline
   & $a$ & $b$ & $c$ & $d$ \\
\hline
S1 & 582.44 & 0.048636 & -15231176.17 & -6.70E-06 \\
S2 & 562.44 & 0.048636 & -15231176.17 & -6.70E-06 \\
\hline
\end{tabular}
\label{Supp_Par3}
\end{table}

\begin{table}[ht]
\centering
\caption{Temperature-dependent interaction parameters for $G^\mathrm{ex}$ according to eqn.~(\ref{Eq_Gex}) for the simulation of the binary system Nb$_2$O$_5$--Ta$_2$O$_5 $.}
\begin{tabular}{ccccc}
\hline
$j$ & $\alpha$-Ta$_2$O$_5$ & $\beta$-Ta$_2$O$_5$ & H-Nb$_2$O$_5$      & Liquid \\
\hline
0   & $5000+3\cdot T$      & $5100+2.9\cdot T$   & $3800+4.6\cdot T$  & $-20000 -5.6\cdot T$ \\
1   & $12000+2.4\cdot T$   & $12000+2.2\cdot T$  & $-9500+4.4\cdot T$ & $3500-2.5\cdot T$ \\
\hline
\end{tabular}
\label{Supp_Par4}
\end{table}

\begin{table}[ht]
\centering
\caption{Temperature-dependent interaction parameters for $G^\mathrm{ex}$ according to eqn.~(\ref{Eq_Gex}) for the simulation of the binary system Nb$_2$O$_5$--V$_2$O$_5 $.}
\begin{tabular}{ccc}
\hline
$j$ & H-Nb$_2$O$_5$        & Liquid\\
\hline
0   & $55000 -34.5\cdot T$ & $-23000 -5.2\cdot T$ \\
1   & $-56000 -7.2\cdot T$ & $-3000 -4.2\cdot T$ \\
\hline
\end{tabular}
\label{Supp_Par5}
\end{table}

\begin{figure}[ht]
\centering
\includegraphics[width=0.65\textwidth]{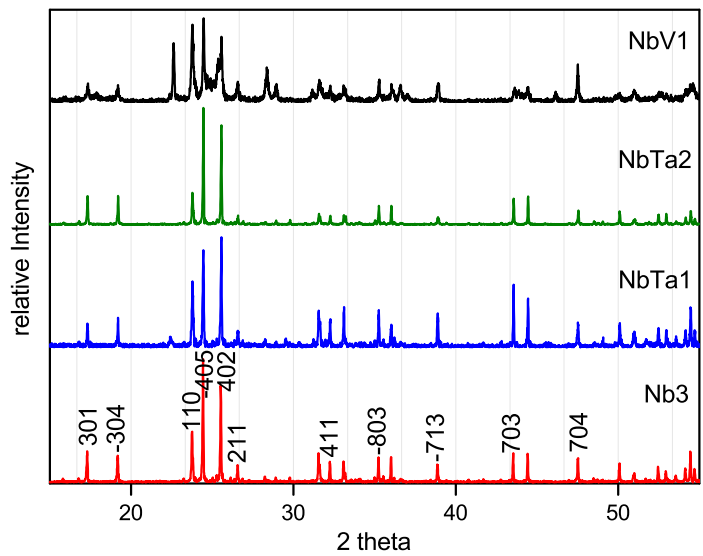}
\caption{X-ray diffraction of the samples NbTa1, NbTa2 and NbV1. All diffractograms show the monoclinic H-modification of niobium(V)-oxide.}
\label{Supp_XRD1}
\end{figure}

\begin{figure}[ht]
\centering
\includegraphics[width=0.65\textwidth]{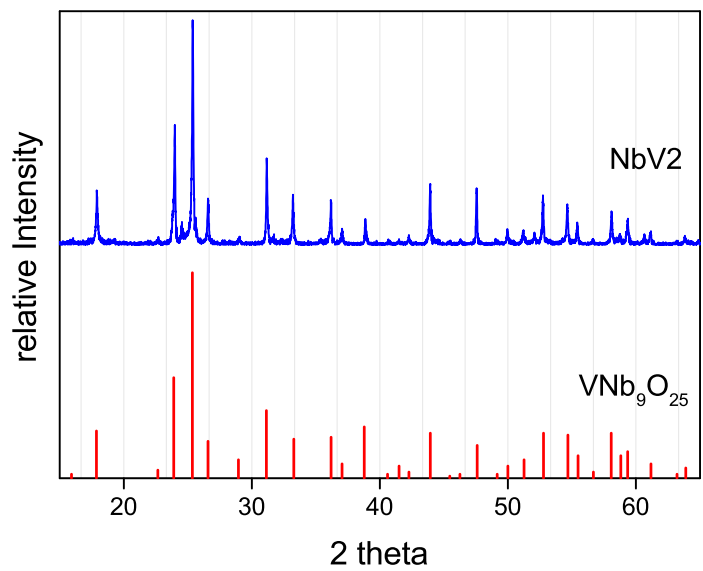}
\caption{X-ray diffractogram of the sample NbV2. All occurring reflexes can be assigned to the intermediate compound VNb$_9$O$_{25}$ \cite{PDF49}.}
\label{Supp_XRD2}
\end{figure}

\end{document}